# Deep Learning for Short-Term Voltage Stability Assessment of Power Systems

**Meng Zhang [1], Jiazheng Li [1], Yang Li [1], Senior Member, IEEE, and Runnan Xu [2]**

[1]School of Electrical Engineering, Northeast Electric Power University, Jilin 132012, China
[2]The Electric and Computer Engineering Department, Illinois Institute of Technology, Chicago, IL 60616, US

Corresponding author: Yang Li (e-mail: liyang@neepu.edu.cn).

This work was supported in part by the Natural Science Foundation of Jilin Province, China under Grant No. 2020122349JC.

**ABSTRACT** To fully learn the latent temporal dependencies from post-disturbance system dynamic trajectories, deep learning is utilized for short-term voltage stability (STVS) assessment of power systems in this paper. First of all, a semi-supervised cluster algorithm is performed to obtain class labels of STVS instances due to the unavailability of reliable quantitative criteria. Secondly, a long short-term memory (LSTM) based assessment model is built through learning the time dependencies from the post-disturbance system dynamics. Finally, the trained assessment model is employed to determine the systems stability status in real time. The test results on the IEEE 39-bus system suggest that the proposed approach manages to assess the stability status of the system accurately and timely. Furthermore, the superiority of the proposed method over traditional shallow learning-based assessment methods has also been proved.

## I. INTRODUCTION

The transmission capacity of power grids is approaching its limit due to electricity market reform and increasing power consumption, which seriously threatens the security and stability operation of power systems. Once the load center areas suffer a large disturbance, considerable dynamic loads tend to restore in several seconds, which could result in short term voltage instability of the power system [1]. Problems caused by the increasing penetration of renewables in different forms such as active distribution networks [2], microgrids [3, 4], and integrated energy systems [5] are driving power systems to potential voltage instabilities because of the renewable uncertainties. Particularly, with the increasing growth of induction motor loads such as air conditioners, short-term voltage stability (STVS) assessment becoming an urgent issue to deal with. As for a power system, STVS is treated as the ability to restore its bus voltage to a normal level within a short period after a fault [1]. According to concrete knowledge of system models, some methods such as energy function [6] and PV plane [7] have been used for STVS. However, it is still a significant and challenging problem to accurately and timely determine whether the power system can maintain the STVS status without the accurate physical models of power systems.

Recently, as the advanced wide-area measurement system infrastructures are widely used in power systems, massive synchronized measurement data is available for data-driven STVS assessment [8-10]. Based on the data taken from phasor measurement units (PMU), great efforts are made to achieve data-driven STVS assessments. Ensemble learning of neural networks with random weights is adopted to assess STVS hierarchically in [9]. Combined with a decision tree (DT), the STVS assessment procedure in [10] is performed on the basis of time series (TSs) shapelet. A light-duty TSs learning machine named wordbook is developed based on DT for high-efficiency feature learning to assess STVS in [11]. Reference [12] uses DT to perform imbalance learning on the STVS assessment. Extreme learning machine (ELM) based approach is proposed to determine the stability status in a hierarchical self-adaptive manner [13]. Reference [14] utilizes a hybrid randomized ensemble model of ELM and random vector functional link network (RVFL), which gives a better assessment accuracy than the single ELM and RVFL. Unfortunately, the above approaches could discard the latent temporal dependencies relationship of complicated power system dynamics. Taking this point into account, a long short-term memory network (LSTM) [15] is introduced to fully extract the time-domain correlation features from TSs data in this paper.

The main contributions of this paper are as follows:

(1) This paper attempts to introduce deep learning for fully capturing the potential temporal dependencies from the post-disturbance power system dynamics. The LSTM network with deep architecture can extract sequential STVS features from PMU data, which is novel in the STVS assessment field.

(2) The simulation results on the IEEE 39-bus system demonstrate that the proposed LSTM-based STVS assessment approach manages to make a more accurate assessment with a faster response speed, comparing with the traditional models based on shallow machine learning such as support vector machine (SVM) and DT.

(3) Besides assessment accuracy, this study carries out statistical tests to comprehensively evaluate the performance of the proposed approach by adopting indicators such as receiver operating characteristic (ROC) curve, area under the curve (AUC), and F1-score.

## II. SEMI-SUPERVISED CLUSTER LEARNING

So far, there still isn't a dependable quantitative criterion to determine the STVS statue of a power system, which makes it not easy to acquire the accurate labels of the instances. Labeling each instance according to domain knowledge inevitably wastes a considerable amount of time, which is not suitable for practical applications. However, it is worth noting that based on explicit cognition of voltage stability, some instances can be easily determined whether can hold STVS or not. For example, for a given power system, all bus voltages of the system are maintained above 0.9pu, then there is no doubt that it can remain stable; if all bus voltages fall below 0.7pu and sustain the low voltage without recovery, it can be identified as unstable noticeably [10]. In this way, the blindness of unsupervised clustering process can also be eliminated [16].

Constraint-partitioning k-means (COP k-means), a representative semi-supervised clustering algorithm is utilized in this work [10]. In COP k-means algorithms, Must/Cannot-link constraints are used to guide the clustering procedure to determine which instances should or should not be in the same cluster.

The main procedures of COP k-means cluster learning are as follows:

Step 1: Input TSs database $\{TS_1,\cdots,TS_s,\cdots,TS_N\}$ and Must/Cannot-link constraints.

Step 2: Initialize the clustering center $TS_{o\beta}$. In terms of the cluster $C_\beta=\{TS_1,\cdots,TS_s,\cdots,TS_N\}$, its center represented as $TS_{o\beta}=\{TS_{o\beta,1},\cdots,TS_{o\beta,i},\cdots,TS_{o\beta,d}\}$ is calculated by:

$$TS_{o\beta,i}=\left\{\left(\frac{1}{N}\sum_{s}^{N}P_{sj}\right)\middle|1\le j\le m, P_{sj}\in TS_s\right\}, 1\le\beta\le k \quad (1)$$

where $k$ means TS database with $k$ clusters, concretely speaking, $k$ is taken as 2; here $TS_s\in TS_s$ and the $j$th point of $TS_s$ is $P_{sj}$; $TS_{o\beta,i}$ is the collection of the sequential elements.

Step 3: Calculate the distance between each instance $TS_s$ and the cluster center $TS_{o\beta}$ according to the following equations and assign $TS_s$ to the cluster whose center is nearer.

$$D(TS_s,TS_{o\beta})=\left(\sum_{i=1}^{d}D(TS_{s,i},TS_{o\beta,i})^2\right)^{1/2} \quad (2)$$

where $d$ is the dimension of each instance $TS_s$; $D(TS_s,TS_{o\beta})$ is the distance defined by the Euclidean norm, taking the one-dimensional TSs data as an example, the specific representation is presented as follows:

$$D(TS_s,TS_{o\beta})=\left(\sum_{j=1}^{m}\left|P_{sj}-P_{o\beta j}\right|^2\right)^{1/2} \quad (3)$$

where $P_{sj}$ and $P_{o\beta j}$ are respectively the $j$th element of the one-dimensional TSs sequence $TS_s$ and $TS_{o\beta}$.

Step 4: If the allocation is not in accord with the Must/Cannot-link constraints, the instance $TS_s$ is reassigned according to the constraint.

Step 5: Iterate between Step 2 and Step 4 until convergence.

Step 6: Output the class labels corresponding to all instances.

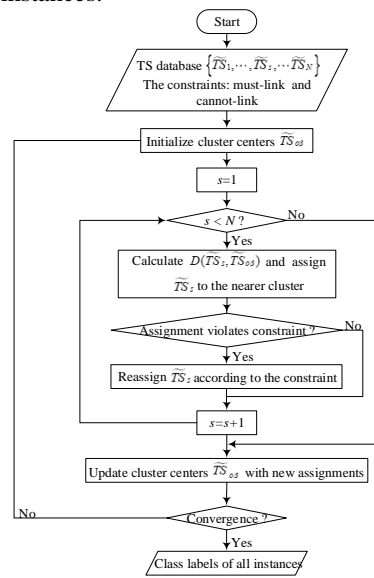

**FIGURE 1.** Flowchart of the cop k-means algorithm.

## III. LSTM FOR STVS

As one of the remarkable recurrent neural networks (RNNs), LSTM has prominent advantages to tackle sequential tasks by capturing the latent temporal dependencies from TSs inputs. The following sections describe the principles of LSTM and the proposed LSTM-based model.

## A. PRINCIPLE OF LSTM

The LSTM network is composed of a stack of cells [17]. Due to the introduction of the memory modules in the LSTM cells, it overcomes the problem of gradient disappearance and explosion existing in the original RNN [17]. The LSTM cell includes three gates: input, forget, and output gates. At each time step, the memory cell can determine how much information is removed or added to the cell state. The input, forget, and output gates are formulated as the following equations [17]:

$$f_t = \sigma(W_f \cdot [h_{-1}, x_t] + b_f) \qquad (4)$$

$$i_t = \sigma(W_i \cdot [h_{-1}, x_t] + b_i) \qquad (5)$$

$$o_t = \sigma(W_o \cdot [h_{-1}, x_t] + b_o) \qquad (6)$$

where $f_t$ is forgotten gate; $i_t$ is the input gate and $o_t$ is the output gate; $W_f$, $W_i$, $W_o$, $W_C$ and $b_f$, $b_i$, $b_o$, $b_C$ are respectively the weight matrixes and the corresponding bias vectors, which is the part that the network needs to learn; $\sigma()$ is the sigmoid function.

The entire transition process of cell state is formulated as follows:

$$C_t = \tanh(W_C[h_{-1}, x_t] + b_C) \qquad (7)$$

$$C_t = f_t * C_{t-1} + i_t * C_t \qquad (8)$$

where $C_t$ is the candidate cell; $*$ is the element-wise product; $C_{t-1}$ respectively represent the cell state at the current time step $t$ and the previous time step $t-1$.

The hidden state $h_t$ is calculated by

$$h_t = o_t * \tanh(C_t) \qquad (9)$$

where tanh is nonlinear activation function. The unique structure which uses memory cells instead of the hidden layer nodes in the LSTM network allows it to extract time-domain correlation feature for a longer period.

## B. LSTM-BASED STVS ASSESSMENT MODEL

Overall, there are mainly five components in the proposed LSTM-based model for STVS assessment, i.e. input, LSTM layer, fully connected (FC) layer, softmax layer, and output, which is shown in Fig. 2.

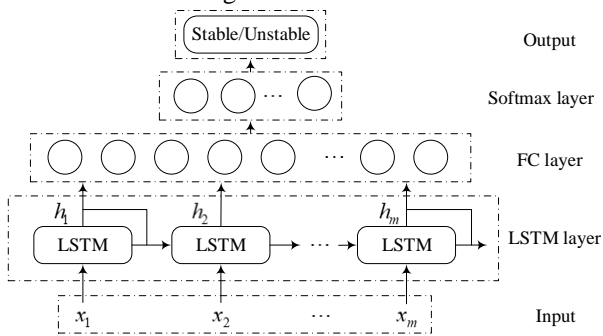

**FIGURE 2. Structure of the LSTM-based model.**

In this assessment model, the input is the post-contingency system dynamics.

The LSTM layer is used to capture the latent temporal dependencies. Benefiting from the internal memories of the LSTM network, the unique gating mechanism is adopted to tackle the TSs inputs in chronological order. In this way, the inputs which are digitized into TSs information with three electrical quantities expressed as $U/P/Q$ including voltage amplitude, the active power, and the reactive power are encoded in the LSTM network parameters.

The FC layer is used for dimensionality reduction and dropout is implemented on this layer to prevent overfitting.

In the softmax layer, the mapping from the hidden features to the final assessment result $\hat{y}$ is realized through the softmax activation function, which is denoted by:

$$\hat{y} = \text{softmax}(W_s \cdot h_t + b_s) \qquad (10)$$

where $h_t$ is the hidden feature; $W_s$, $b_s$ are the weights and biases which are needed to be tuned during the offline training process.

The final output $\hat{y}$ reflects the stability status of the power system.

## IV. PROPOSED APPROACH

The proposed approach is mainly composed of three stages. Firstly, for a given power system, time-domain simulations (TDSs) are performed to generate a TSs database, and the class labels of all instances are acquired through the COP k-means algorithm. Secondly, the offline training process is performed on the established LSTM-based model. Finally, the LSTM-based model is used for online applications. The flowchart of the proposed approach is shown in Fig. 3, the details of each stage are shown in each section.

## A. DATASET GENERATION

For a given power system, TDSs are performed for all kinds of possible post-disturbance system behaviors to generate a comprehensive TSs database. Concerning various system operating conditions, the occurrence of contingencies is simulated by setting the different fault types, fault locations, and fault clearing time with different possible operating parameters such as different proportions of dynamic load during TDSs. As for each instance of the TSs database, the $U/P/Q$ electrical quantities are collected by post-contingency PMU measurements. The database $D_{TS}$ is described as follows:

$$D_{TS} = \{TS_1, TS_2, \cdots, TS_s, \cdots TS_N\}, s \in [1, \cdots, N] \qquad (11)$$

here $N$ instances are generated by TDSs and then integrated into a TSs database $D_{TS}$ for STVS assessment; $TS_s$ represents the TSs collection of the $s$th instance, which is expressed as:

$$TS_s = \begin{cases} TS_{s,1}, \cdots, TS_{s,L} & , \text{ for } U \\ TS_{s,L+1}, \cdots, TS_{s,2L} & , \text{ for } P \\ TS_{s,2L+1}, \cdots, TS_{s,3L} & , \text{ for } Q \end{cases} \qquad (12)$$

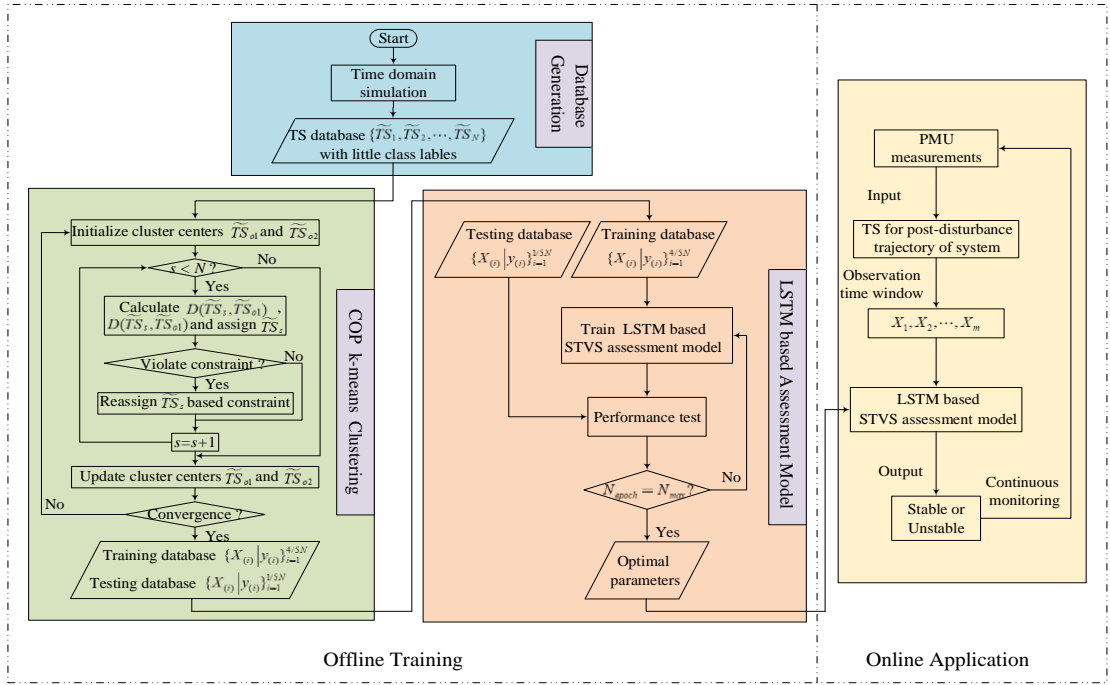

**FIGURE 3.** Flowchart of the proposed approach.

where $L$ denotes the total numbers of the buses; $d$ represents the dimensions of the multivariate $TS_s$ and here $d = 3L$. It means that if there are $L$ buses in a given power system, the dimension of the multivariate TSs instance is $3L$.

For a post-contingency observation time window (OTW) with length $T$, the sampling time is set to $\Delta t$. Hence, the length of the TS of each electrical quantity is denoted as $m$, would be $m = T/\Delta t$. At the same time, all instances in the TSs database are normalized to eliminate the influence of $U/P/Q$ in different ranges.

### B. OFFLINE TRAINING STAGE

At the offline training stage, the LSTM-based assessment model is trained to excavate the potential temporal dependencies. In the STVS sequential task, the LSTM-based model takes TSs with three electrical quantities $U/P/Q$ and the class label $y$ obtained by COP k-means algorithm together as the input. Consequently, the training dataset $\{X_{(i)}\big|y_{(i)}\}_{i=1}^{4/5 N}$ can be obtained and each sample of the dataset is denoted by:

$$X_{(i)} = \{X_{(i)1}, X_{(i)2}, \cdots, X_{(i)t}, \cdots, X_{(i)m}\big|y_{(i)}\} \tag{13}$$

where $X_{(i)t}$ characterizes the system dynamics at time step $t$, the elements in $X_{(i)t}$ is as follows:

$$X_{(i)t} = \{U_{t,1}, \cdots U_{t,L}, P_{t,1} \cdots P_{t,L}, Q_{t,1} \cdots Q_{t,L}\} \tag{14}$$

In the offline training stage, the parameters of the LSTM network including $W$ and $b$ are continuously fine-tuned. Stochastic gradient descent is optimized by Adam to find optimal parameters of the LSTM-based model. And the $L2$ norm applied to the Euclidean distance serves as the loss function of the model [18]. Based on the loss function, the network parameters of LSTM are continuously fine-tuning during the iteration process [19].

### C. ONLINE APPLICATION STAGE

At the online application stage, the trained model with the optimal parameters through performance tests is employed for online real-time STVS assessment. Via PMU measurements or other data acquisition devices, the dynamics of the system can be monitored in real-time. Meanwhile, this system dynamics information can be quantified as TSs and input into the trained model for a dependable assessment result to deal with the possible unexpected contingency. Attributing to the significant learning efficiency of the LSTM-based model, reliable assessment results can be obtained immediately. In this way, based on the assessment results which reflect the stability status, remedial control measures will be started up in time to avoid possibly serious cascading collapse.

### D. EVALUATION INDEX

In addition to assessment accuracy, statistical indicators including AUC and F1-score are utilized to comprehensively evaluate the proposed approach in this paper [20]. And these indicators are all defined according to the confusion matrix which is shown in Table I.

TABLE I
CONFUSION MATRIX

| Confusion matrix | Stable(Actual) | Unstable(Actual) |
|---|---|---|
| Stable (Predicted) | *TP* | *FP* |
| Unstable (Predicted) | *FN* | *TN* |

In the confusion matrix of Table I, as for a stable sample, if it is classified as stable, it is regarded as true positive (*TP*) and if it is assessed as unstable, it is regarded as false negative (*FP*). As for an unstable sample, if it is assessed as unstable, it is regarded as true negative (*TN*), otherwise, it is regarded as false positive (*FN*).

### 1) Accuracy

Accuracy is widely used to evaluate the performance of the STVS assessment approaches [7-14].

$$\text{Accuracy} = \frac{TP + TN}{TP + FP + FN + TN} \quad (15)$$

Accuracy is the proportion of cases that are correctly predicted in the assessment process, which provides an overall performance evaluation of the STVS assessment approach.

### 2) ROC curve

ROC curve is a basic tool for analyzing and evaluating classification models [20]. In this paper, the ROC curve is mainly used to evaluate the misdetection of classifiers. As for the two-dimensional ROC curve, the X-axis and Y-axis respectively represent the false positive rate (FPR) and the true positive rate (TPR), denoting by:

$$\text{TPR} = \frac{TP}{TP + FN} \quad (16)$$

$$\text{FPR} = \frac{FP}{FP + TN} \quad (17)$$

In the ROC curve, the classifiers corresponding to the upper left-hand side points with superior performance than those corresponding to other points. They classify nearly all stable samples correctly, meanwhile, they often have low misclassification rates of the unstable samples, which is critical for STVS assessment.

### 3) AUC

AUC is the area under the ROC curve [20]. Its values of the corresponding ROC curves help to distinguish the better ROC curve at the point of intersections. A higher value of AUC implies a better performance of the classification model.

### 4) F1-score

F1-score which is given in (18) is commonly used to evaluate the binary classifier [21]. It is the weighted average of TPR and FPR. The value of the F1-score is within the interval [0, 1]. The higher value of the F1-score, the better performance of the classifier

$$\text{F1-score} = 2 \cdot \frac{TPR \cdot FPR}{TPR + FPR} \quad (18)$$

## V. CASE STUDY

The test results on the IEEE 39-bus test system which is widely-used for SVTS assessment verify the superiority of the proposed approach in assessment performance [9,13,14], comparing with other state-of-the-art approaches based on DT and SVM. All experiments have been implemented via the Google TensorFlow 1.14.0. NVIDIA and CUDA 10.0 are utilized for the usage of GPU computation power. Note that PMU data is simulated through detailed TDSs by using the commercial software PSD-BPA.

### A. DATABASE GENERATION

Considering the different contingencies under various operation conditions, TDSs are performed to approximate all kinds of post-disturbance system behaviors. (i) As for the total load demand, it is set respectively 80%, 100%, and 120% of the base level; (ii) the proportions of motor load are respectively adjusted to 70%, 80%, and 90%; (iii) Starting from the terminal of one transmission line, three-phase short-circuits fault is separately imposed on 0, 20%, 40%, 60%, and 80% of the whole length; (iv) when short-circuit fault occurred on the transmission line at the time of 0.1s, it will be cleared within 0.15s to 0.2s. Based on the above settings, a comprehensive database contained 1200 samples is built, meanwhile, the training and testing dataset can be obtained by dividing it according to a ratio of 4:1 [9, 10].

### B. PERFORMANCE TEST ANALYSIS

In this section, the hyper-parameters of LSTM are detailed in Table II.

TABLE II
HYPER-PARAMENTS SETTING OF LSTM.

| Hyper-parameters | Values |
|---|---|
| Learning rate | 0.0001 |
| Dropout | 0.25 |
| LSTM cell size | 256 |
| Batch size | 64 |
| Epoch | 200 |

In the LSTM network adopted in this paper, the number of LSTM cells is 256, the learning rate of the Adam optimizer is $1 \times 10^{-4}$, and 64 batch sizes are used to maximize the usage of GPU resources. Moreover, to prevent overfitting, a 25% dropout for the LSTM and the FC layer is implemented [22]. The LSTM-based model is trained for 200 epochs.

On the basis of the above hyper-parameter settings, the performance tests are carried out in the OTW with different lengths 0.03s, 0.06s, 0.09s, and 0.12s. And the performance test results of accuracies are shown in Fig. 4.

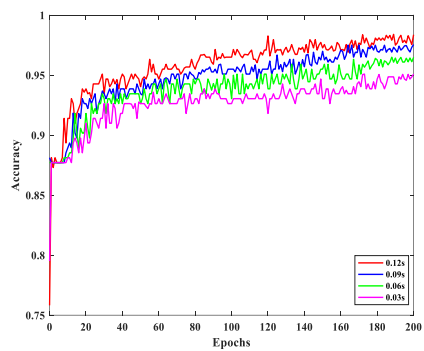

**FIGURE 4.** Accuracies of the proposed approach at OTW with different lengths.

As shown in Fig. 4, it can be noted that the rising curves imply that network parameters of the LSTM-based model are gradually optimized. As the length of OTW increasing, the overall assessment accuracies are significantly enhanced, which proves its outstanding advantage of LSTM in learning from the temporal dependencies. Taking the cost of misclassification into account, the ROC curve is adopted in this paper to confirm the reliability of the proposed method and the ROC curves within the OTW with various lengths are as shown in Fig. 5.

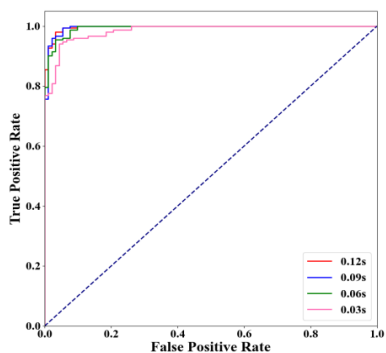

**FIGURE 5.** ROC curves of the proposed approach at OTW with different lengths.

It can be observed that the four curves are all distributed on the upper left-hand side, which implies that the total number of stable cases which correctly predicted as stable is considerably much, meanwhile, the total number of unstable cases which are misclassified as stable is extremely few. Since the unstable samples can be detected with a high probability, the control devices can be started up immediately to ensure the STVS of the power system, which proves the reliability of the assessment result. What's more, considering the essence of the assessment model is a binary classifier, F1-score is employed for the performance analysis.

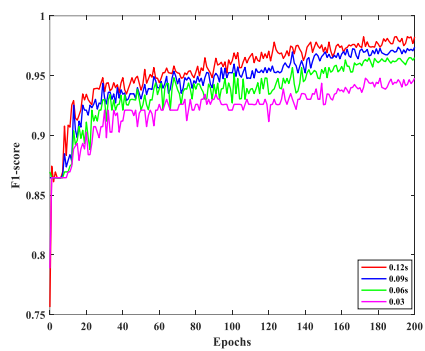

**FIGURE 6.** F1-scores of the proposed approach at OTW with different lengths.

It can be discovered from Fig. 6 that the overall distribution of F1-scores is maintained at a high level above 0.95, which implies a high-quality classification performance of the proposed STVS assessment approach.

In summaries, according to the above performance test results, the proposed LSTM-based model shows its great advantages in assessment accuracy, reliability, and classification performance while handling the sequential STVS tasks.

## C. COMPARISON TEST ANALYSIS
The comparison test analysis of the proposed approach is carried out with the comparison of traditional shallow machine learning-based models, here, two representative algorithms DT and SVM are taking as examples. Taking all the evaluation indicators, i.e. accuracy, AUC, and F1-score into account, the comparison test results are listed in Table III.

TABLE III
COMPARISONS OF DIFFERENT APPROACHES ABOUT THE LENGTH OF OTW, ACCURACY, F1-SCORE, AND AUC.

| Algorithms | OTW(cycles) | Accuracy(%) | F1-score(%) | AUC |
|---|---|---|---|---|
| LSTM | 3 | 95.08 | 0.9479 | 0.9855 |
| | 6 | 96.72 | 0.9651 | 0.9936 |
| | 9 | 97.54 | 0.9738 | 0.9954 |
| | 12 | 98.36 | 0.9825 | 0.9963 |
| DT | 3 | 92.21 | 0.9183 | 0.9392 |
| | 6 | 93.03 | 0.9269 | 0.9459 |
| | 9 | 93.44 | 0.9313 | 0.9482 |
| | 12 | 93.44 | 0.9313 | 0.9482 |
| SVM | 3 | 87.70 | 0.8646 | 0.9509 |
| | 6 | 87.70 | 0.8646 | 0.9672 |
| | 9 | 87.70 | 0.8646 | 0.9757 |
| | 12 | 87.70 | 0.8646 | 0.9781 |

It can be seen that within the OTW of the four lengths, the accuracies of the LSTM-based model are much higher than those of DT and SVM-based models, and so do those in F1-score and AUC. Meanwhile, it is worth noting that, the accuracy of the LSTM-based model is 95.08% at the OTW with length 3 cycles, which is respectively 1.72% and 7.76% higher than that of DT and SVM-based models at the OTW with length 12 cycles. In other words, the proposed approach can achieve higher accuracy with a faster response speed, which fully certifies the superiority of the proposed approach. And considering the cost of misclassification, the comparison test results are obtained as shown in Fig. 7.

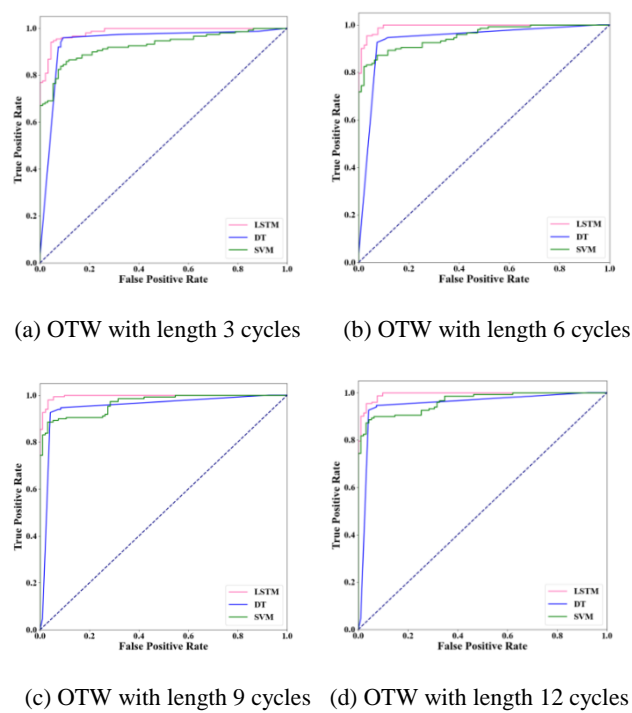

(a) OTW with length 3 cycles    (b) OTW with length 6 cycles

(c) OTW with length 9 cycles    (d) OTW with length 12 cycles

**FIGURE 7.** ROC curve comparison of different approaches at OTW with different lengths.

It can be observed that the ROC curves of the LSTM-based model are always distributed above both the DT and SVM-based models no matter what the length of OTW. It illustrates that the LSTM-based model misclassifies the unstable cases as stable much less when the amount of actual stable cases that are correctly predicted are equal. Furthermore, it indicates that the proposed approach manages to obtain more reliable assessment results comparing with the DT and the SVM-based models.

## VI. CONCLUSION

In order to fully capture the latent temporal dependencies from post-disturbance dynamic trajectories, a novel data-driven STVS assessment approach is proposed for power systems in this paper by introducing deep learning. Through analysis and discussion, the main conclusions can be drawn as follows:

(1) Unlike the existing state-of-the-art assessment approaches, the LSTM-based STVS approach in this paper learns from the temporal data dependencies, which improves assessment accuracy and significantly reduces the length of OTW.

(2) Besides accuracy, considering AUC and F1-score, the simulation results on IEEE 39-bus system comprehensively validate the effectiveness of the proposed approach.

(3) Furthermore, the superiority of the proposed deep learning-based assessment approach over traditional shallow learning-based approaches has also been verified.

The future work is dedicated to improving the adaptability of data-driven assessment approaches considering the missing PMU measurements and class imbalance. Another interesting topic is to study the STVS of power systems with new elements such as high voltage direct current (VSC-HVDC) [23, 24] and combined heat and power plants [25, 26].